\documentclass[smallextended]{MYsvjour3}       
\smartqed  
\usepackage{bm,amsfonts,amssymb,amsmath,graphicx}
\newcommand{\OMEGA}{\mbox{\boldmath${\omega}$}}
\newcommand{\Area}{\mbox{\boldmath${A}$}}
 \journalname{Gen Relativ Gravit (2012) 44: 1713-1723}
 \doi{DOI 10.1007/s10714-012-1361-4}
 \http{http://www.springerlink.com/content/2q3081435g3568v1/}

\begin{document}

\title{General relativistic Sagnac formula revised}


\author{Paolo~Maraner \and Jean-Pierre~Zendri}


\institute{Paolo~Maraner  \at
              School~of~Economics~and~Management,~Free~University~of~Bozen-Bolzano,~via~Sernesi~1,~39100~Bolzano,~Italy \\
              \email{pmaraner@unibz.it}           
           \and
           Jean-Pierre~Zendri \at
              INFN sezione di Padova,~via~Marzolo~8,~35131~Padova,~Italy\\
              \email{Jean-Pierre.Zendri@lnl.infn.it}
}

\date{}

\maketitle

\begin{abstract}
The Sagnac effect is a time or phase shift observed between two beams of light traveling in opposite directions in a rotating interferometer. We show that the standard description of this effect within the  framework of general relativity misses the effect of deflection of light due to rotational inertial forces. We derive the necessary modification and demonstrate it through a detailed analysis of the square Sagnac interferometer rotating about its symmetry axis in Minkowski space-time. The role of the time shift in a Sagnac interferometer
in the synchronization procedure of remote clocks  as well as its analogy with the Aharanov-Bohm effect are revised.
\end{abstract}

\section{Introduction}
\label{intro}
\begin{figure}[t]
 \centering
 \includegraphics[width=7.0cm]{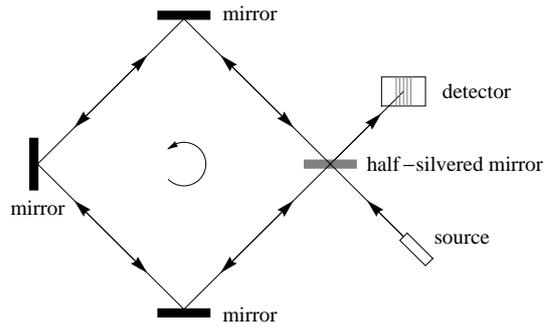}
\caption{A square Sagnac interferometer rotating about its symmetry axis in an inertial frame of reference.}
 \label{inertial}
\end{figure}

\begin{figure}[t]\label{2}
 \centering
 \includegraphics[width=7.0cm]{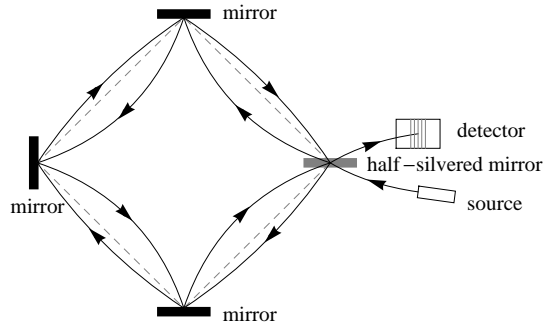}
\caption{A square Sagnac interferometer at rest in a non-inertial frame of reference rotating about the
symmetry axis of the interferometer. The natural rotation parameter $\omega\bar{r}/c$ is set to about 0.3.}
 \label{non inertial}
\end{figure}

Consider a beam of monochromatic light split by a half-silvered mirror in two beams, and those two beams
directed in a close path around a set of mirrors in opposite directions (see Fig.\ \ref{inertial}). Subsequently observe the
interference pattern of the beams emerging out of the half-silvered mirror.
If the apparatus is at rest with respect to a local reference frame of inertia, the beams will travel the same
distance at the same speed, so they will arrive at the end point simultaneously.
If the apparatus is set in rotation, they will no longer arrive simultaneously.
The leading contribution to the arrival time shift $\Delta t$ is given by
\begin{equation}
c\Delta t\approx\frac{4}{c}\Area\cdot\OMEGA,
\label{S1913}
\end{equation}
with $\Area$ the vector area embraced by the interferometer, $\OMEGA$ the angular speed of rotation
and $c$ the speed of light.
This was predicted and observed by Sagnac in 1913 \cite{Sag13,Post67,Chow85,ABS94}.\footnote{An analogous effect
is predicted \cite{Pag75,Ana77,DY79} and observed \cite{WSC79,RKW91,HN93,{STK11}} for matter beams. In this paper we focus on the optical
Sagnac effect. Analogies and differences with the Sagnac effect for matter beams are discussed in section \ref{sec:3}.}
The interpretation of the Sagnac effect, as well as the computation of relativistic corrections to (\ref{S1913}), have
been the subject of endless controversies since.\footnote{Leaving aside claims that the Sagnac effect
entails a conflict with the principles of relativity, it remains the fact that there is still no full agreement on its
explanation in a general context. Corrections to the Sagnac formula beyond those linear in $\omega$ are still to be univocally
determined. See e.g.\ the remarks on p.\ 145 of \cite{HN93}, and subsection 2.3 of
\cite{Ste97}. See also \cite{Malykin00} and \cite{KBFS09}.}
Nonetheless, compasses of inertia based on this effect
are today capable of detecting rotation rates as small as $10^{-12}$ radiant per second \cite{Wet11}.
Its applications are  in commercial navigation systems,
in Earth rotation monitoring \cite{Wet11,SKVSS09}
and might soon provide a direct Earth-bound measure of Lense-Thirring frame dragging \cite{SSB03,DiVetA10,{B&al11}}.
Admittedly relativistic corrections to (\ref{S1913}) are too small to be of practical interest in present
applications. Their correct evaluation is important in planing future experiments and from the theoretical point of view however.
On the theoretical side the effect is involved in the synchronization procedure of remote clocks \cite{AVS98,Min03},
provides a gravitational analogue of the Aharonov-Bohm effect \cite{AsMa75,Sak80,Ana81,HN90,RR03}
and a general characterization of rotations in general relativity \cite{AsMa75}.
Up to now its description within the framework of general relativity
is given as follows: in a stationary space-time with coordinates $x=(ct,\vec{x})$ and
line element $ds^2=g_{\mu\nu}(\vec{x})dx^\mu dx^\nu$, the coordinate-time shift $\Delta t$ is given by the formula
\begin{equation}
c\Delta t=-2\oint_{C}\frac{g_{0i}}{g_{00}}dx^i,
\label{S}
\end{equation}
with $C$ the closed path traveled by light in the interferometer;\footnote{The interferometer is at rest in the
given coordinates frame. Metric entries do not depend on the time-coordinate $t$. Greek indices $\mu,\nu,...$ range form 0 to 3
while Latin ones $i,j,...$ from 1 to 3. Our signature convention is $-,+,+,+$.}
the proper-time shift $\Delta\tau$, which is the quantity directly related to the phase shift measured at the detector,
is then obtained as
\begin{equation}
\Delta\tau=\sqrt{\left|g_{00}(\vec{x}_{det.})\right|}\Delta t,\hskip0.5cm
\label{Spt}
\end{equation}
with $g_{00}(\vec{x}_{det.})$  the $00$ metric entry evaluated at detector's position.
To the leading order (\ref{Spt}) reproduces (\ref{S1913}), plus relativistic corrections.
The proof of (\ref{S}) essentially parallels the computation of the coordinate-time delay in the
synchronization procedure of a clock transported along a closed path.\footnote{The
standard reference is \cite{LL}, where the formula is obtained in relation to the
synchronization of clocks transported along closed paths, without mention to the Sagnac effect.
With explicit mention to the Sagnac effect, the formula is re-derived with different notations
and/or particular assumptions in a number of papers. For a recent re-derivation see e.g.\ \cite{KBFS09}.
}
More sophisticated, but substantially equivalent proofs are given in \cite{AsMa75,Ana81}.\\
In this paper we would like to point out that equation (2) only provides the correct
leading order to the time delay in a Saganc interferometer, while it is not suitable for the computation of relativistic corrections.
This is because of the implicit assumption that the two beams propagating in opposite directions  in a
rotating Sagnac interferometer follow an identical path.
Nevertheless, in a rotating frame of reference, null geodesics connecting different space-time
points in opposite directions are in general not coincident (see Fig.\ \ref{non inertial}).\footnote{This
effect is automatically taken into account in the general formalism describing the split
and propagation of light in relativity, developed by S.\ L.\ Ba\.{z}a\'{n}ski
\cite{Baz98-99}.} This  implies the
inadequacy of (\ref{S}) in computing relativistic corrections and  the need for its revision.
In order to obtain an appropriate general relativistic Sagnac formula, we go over
the derivation of (\ref{S}) and introduce the necessary changes. The
outcome is considerably less elegant than (\ref{S}) and no longer describes the time delay in a Sagnac interferometer
as the circulation of a vector field. It nevertheless provides the right mean of performing the computation.
To demonstrate this in a specific case, we focus on the square Sagnac interferometer rotating about its axis in
Minkowski space-time. We compute the exact time shift in an inertial frame of reference and show disagreement
with the prediction of (\ref{S}) and agreement with the prediction of the revised formula.
In order to estimate relativistic corrections in a realistic case, we then consider the square interferometer
rotating around a point different from its center. Finally, we discuss some implications of the revised formula.

\section{Derivation of Sagnac formula}
\label{sec:1}
Parameterizing the null geodesic followed by either the clockwise or counterclockwise beam
as $\gamma(z)=(ct(z),\vec{x}(z))$ with $z\in\left[0,\bar{z}\right]$, we have that
\[
c^2g_{00}\left(\frac{dt}{dz}\right)^2+2cg_{0i}\frac{dx^i}{dz}\frac{dt}{dz}+g_{ij}\frac{dx^i}{dz}\frac{dx^j}{dz}=0.
\label{ds^2=0}
\]
The metric entries $g_{00}$, $g_{0i}$ and $g_{ij}$ are evaluated along the null geodesic $\gamma(z)$ or equivalently,
given  the coordinate-time independence of the metric, along the spatial path $C(z)=(0,\vec{x}(z))$ followed
by the beam in the interferometer.
The solutions of the above quadratic equation
read
\[
c\left(\frac{dt}{d z}\right)_{\pm}=-\frac{g_{0i}}{g_{00}}\frac{dx^i}{dz}\pm
\sqrt{\frac{\gamma_{ij}}{|g_{00}|}\frac{dx^i}{dz}\frac{dx^j}{dz}},
\label{dt_pm}
\]
where $\gamma_{ij}=g_{ij}-g_{0i}g_{0j}/g_{00}$ is the three-dimensional positive definite spatial metric induced
by the space-time metric \cite{LL}.
As usual, we identify  the positive solution $(dt/dz)_{+}$ with the counterclockwise beam and the negative solution
$(dt/dz)_{-}$  with the clockwise beam. At this point, however, we have to distinguish between the spatial path
$C_{\circlearrowleft}(z)$ followed by the counterclockwise beam and the spatial path $C_{\circlearrowright}(z)$
followed by the clockwise one. The total coordinate-time $t_{\circlearrowleft}$ traveled by the
counterclockwise beam is then given by
\[
ct_{\circlearrowleft}=c\int_{0}^{\bar{z}}\left(\frac{dt}{dz}\right)_{+}dz=
-\oint_{C_{\circlearrowleft}}\left(\frac{g_{0i}}{g_{00}}dx^i-\frac{1}{\sqrt{|g_{00}|}}dl\right),
\]
where $dl=\sqrt{\gamma_{ij}dx^idx^j}$ is the induced spatial line element.
The total coordinate-time $t_{\circlearrowright}$ traveled by the clockwise beam reads instead
\[
ct_{\circlearrowright}=c\int_{\bar{z}}^{0}\left(\frac{dt}{dz}\right)_{-}dz=
-c\int_{0}^{\bar{z}}\left(\frac{dt}{dz}\right)_{-}dz=\oint_{C_{\circlearrowright}}\left(\frac{g_{0i}}{g_{00}}dx^i+\frac{1}{\sqrt{|g_{00}|}}dl\right),
\]
with analogous notations. Correspondingly, the coordinate-time shift
$\Delta t=t_{\circlearrowleft}-t_{\circlearrowright}$ is obtained as
\begin{eqnarray}
c\Delta t=
-\oint_{C_{\circlearrowleft}}\frac{g_{0i}}{g_{00}}dx^i
-\oint_{C_{\circlearrowright}}\frac{g_{0i}}{g_{00}}dx^i
+\oint_{C_{\circlearrowleft}}\frac{1}{\sqrt{|g_{00}|}}dl
-\oint_{C_{\circlearrowright}}\frac{1}{\sqrt{|g_{00}|}}dl.
\label{Sr}
\end{eqnarray}
Given the differences in the integration paths, the first two integrals  no longer sum up as in (\ref{S})
and the third and fourth ones no longer cancel. The proper time shift $\Delta\tau$ measured at the detector
is again obtained by multiplying $\Delta t$ by $\sqrt{\left|g_{00}(\vec{x}_{det.})\right|}$ as in (\ref{Spt}).

\section{Square Sagnac interferometer}
\label{sec:2}
In order to demonstrate in a specific case the need for the revision of (\ref{S}),
we proceed by computing the time shift for the square Sagnac interferometer displayed in figures \ref{inertial}
and \ref{non inertial}, rotating about its symmetry axis with angular speed $\omega$ in Minkowski space-time.
This allows us to perform the computation in two distinct and independent ways.
First, in an inertial frame of reference where the interferometer is rotating.
This approach does not employ the formalism of general relativity and provides us with the exact result to an arbitrary accuracy.
Then, we evaluate the same time shift in a non-inertial frame of reference
where the interferometer is at rest by means of (\ref{S}) and (\ref{Sr}).
The comparison with the exact result shows disagreement with the prediction of (\ref{S}) and agreement with the one of (\ref{Sr}),
also displaying the order of magnitude of the various contributions.

\paragraph{Inertial frame of reference.}
In Figure \ref{inertial} consider a light pulse emitted at $t=0$  from the
half-silvered mirror toward the upper mirror. In a reference frame centered in the interferometer, with left-right
oriented horizontal $x$-axis and bottom-up oriented vertical $y$-axis,
we can take the upper mirror to move according to
\[
x(t)=\mp\bar{r}\sin\omega t, \hskip0.4cm  y(t)=\bar{r}\cos\omega t,
\]
where $\bar{r}$ is the radius of the interferometer and $\mp$ refer to counterclockwise, resp.\ clockwise, rotation.
Correspondingly the light pulse propagates according to
\begin{equation}\label{lom}
x(t)=\bar{r}+ct\cos\alpha, \hskip0.4cm y(t)=ct\sin\alpha,
\end{equation}
with $c$ the speed of light and $\alpha$ the angle of emission measured in counterclockwise versus with respect to the
positive direction of the $x$-axis. After introducing the dimensionless quantities
\[
z=\frac{c t}{\bar{r}} \hskip0.3cm \mbox{and } \hskip0.3cm \epsilon=\frac{\omega\bar{r}}{c},
\]
the match conditions at time $\bar{t}$ conveniently write as
\begin{equation}\label{match}
\left\{
\begin{array}{l}
  \bar{z}\cos\alpha=-(1\pm\sin\epsilon\bar{z}) \\
  \bar{z}\sin\alpha=\cos\epsilon\bar{z}
\end{array}
\right.,
\end{equation}
with $\bar{z}=c \bar{t}/\bar{r}$. Squaring both equations and adding term by term, we have
\begin{equation}\label{sqeq}
\bar{z}^2=2\left(1\pm\sin\epsilon \bar{z}\right).
\end{equation}
This can be solved by a series expansion in the natural parameter $\epsilon$. After rescaling,  we obtain the times
$\bar{t}_{\circlearrowleft}$ and $\bar{t}_{\circlearrowright}$  needed for a light pulse to travel from one mirror
to the next in the counterclockwise, resp.\ clockwise, direction as
\begin{equation}\label{ub}
\bar{t}_{\circlearrowleft,\circlearrowright}=\frac{\bar{r}}{c}\left(\sqrt{2}\pm\epsilon+\frac{\sqrt{2}}{4}\epsilon^2
\mp\frac{1}{3}\epsilon^3-\frac{17\sqrt{2}}{32}\epsilon^4+ ...\right).
\end{equation}
For later use we also evaluate the emission angles $\alpha_{\circlearrowleft}$ and $\alpha_{\circlearrowright}$ for the
counterclockwise, resp.\ clockwise, beam, by substituting $\bar{t}_{\circlearrowleft}$, resp.\
$\bar{t}_{\circlearrowright}$, back in to (\ref{match})
\begin{equation}\label{angle}
\alpha_{\circlearrowleft,\circlearrowright}=\frac{3}{4}\pi\pm\frac{\sqrt{2}}{2}\epsilon+\frac{1}{2}\epsilon^2
\pm\frac{\sqrt{2}}{8}\epsilon^3-\frac{1}{6}\epsilon^4+ ...\, .
\end{equation}
Finally, we obtain the coordinate-time delay $\Delta t$ for the square interferometer
multiplying by four the difference $\bar{t}_{\circlearrowleft}-\bar{t}_{\circlearrowright}$
\begin{equation}\label{DtS}
c\Delta t=\frac{8\bar{r}^2\omega}{c}
\left[
1-\frac{1}{3}\left(\frac{\omega\bar{r}}{c}\right)^2-\frac{19}{30}\left(\frac{\omega\bar{r}}{c}\right)^4+...
\right].
\end{equation}
The corresponding proper-time delay is obtained by further multiplying $\Delta t$ by the relativistic factor
$\sqrt{1-v^2_{det.}/c^2}$, with $v_{det.}\equiv\omega r_{det.}$ the constant speed of the detector and $r_{det.}$
its radial position. By assuming $r_{det.}\simeq\bar{r}$ we obtain
\begin{equation}\label{DtauS}
c \Delta\tau=\frac{8\bar{r}^2\omega}{c}
\left[
1-\frac{5}{6}\left(\frac{\omega\bar{r}}{c}\right)^2-\frac{71}{120}\left(\frac{\omega\bar{r}}{c}\right)^4+...
\right].
\end{equation}
The leading term is the original Sagnac term, given by four times the area enclosed by the light path, times the angular speed $\omega$,
divided by the squared speed of light.
The remaining terms are tiny corrections, relevant from a theoretical point of view but
hardly detectable experimentally. As a concrete example,
let us imagine to place the large ring laser G in Wettzell on a turntable. This interferometer has
now reached a theoretical sensitivity $\Delta\omega\simeq 10^{-12}\textrm{rad}/\textrm{s}$ and $\bar{r}\simeq 3\textrm{m}$ \cite{Wet11}.
Setting G in rotation at the realistic angular speed of $\omega\simeq 1 \textrm{rad}/\textrm{s}$ one obtains
a relative sensitivity ${\Delta\omega}/{\omega}\simeq 10^{-12}$. This is some four order of magnitude below the relative sensitivity ${\Delta\omega}/{\omega}=\frac{5}{6}\left(\frac{\omega \bar{r}}{c}\right)^2 \simeq10^{-16}$, needed for the detection of relativistic
corrections.
Relativistic effects become visible at an angular speed $\omega\simeq 30 \textrm{rad}/\textrm{s}$.

\paragraph{Non-inertial frame of reference.}
We now repeat the computation in the  framework of general
relativity first by means of (\ref{S}) and then by means of (\ref{Sr}). We start by transforming to a rotating frame of reference
where the interferometer is at rest \cite{Rin}
\begin{equation}\label{transformation}
t\rightarrow t, \hskip0.4cm r\rightarrow r, \hskip0.4cm \phi\rightarrow\phi-\omega t.
\end{equation}
In the new coordinates the line element reads
\begin{equation*}\label{ds2}
ds^2=-\left(1-\frac{\omega^2r^2}{c^2}\right)c^2dt^2+dr^2+r^2d\phi^2 + 2\omega r^2 dt d\phi,
\end{equation*}
whence the metric entries $g_{tt}$, $g_{rr}$, $g_{\phi\phi}$ and $g_{t\phi}$. We then apply the formulas
\begin{description}
\item[(\ref{S}):] In order to apply (\ref{S}), we parameterize the Euclidean straight segment connecting the half-silvered
mirror with the upper mirror as
\begin{equation*}\label{p1}
r(w)=\bar{r}\frac{\sqrt{1+w^2}}{1+w},\hskip0.4cm \phi(w)=\arctan\left(w\right),
\end{equation*}
with $w\in[0,+\infty)$. Substitution in
$g_{t\phi}/g_{tt}\cdot d\phi/dw$,
expansion in $\epsilon$,
integration in $dw$ form $0$ to $+\infty$ and multiplication by $4$, leads to the coordinate-time shift
\begin{equation}
c\Delta t=-2\oint_{C}\frac{g_{t\phi}}{g_{tt}}d\phi
=\frac{8\bar{r}^2\omega}{c}
\left[
1+\frac{2}{3}\left(\frac{\omega\bar{r}}{c}\right)^2+\frac{7}{15}\left(\frac{\omega\bar{r}}{c}\right)^4+...
\right].
\end{equation}
As expected, the expansion fails to reproduce (\ref{DtS}) already at the next-to-leading
order because light pulses do not propagate along straight segments in the co-moving frame of reference.
\item[(\ref{Sr}):] In order to apply (\ref{Sr}), we compute the paths effectively followed by the counterclockwise and clockwise
beams (see Fig.\ \ref{non inertial}) by solving the corresponding geodesics equations or, equivalently,
by  transforming (\ref{lom}) to rotating coordinates by means of (\ref{transformation}). We obtain
\[
r(z)=\bar{r}\sqrt{1+2z\cos\alpha+z^2},\hskip0.4cm
\phi(z)=\arctan\left(\frac{z\sin\alpha}{1+z\cos\alpha}\right)\mp\epsilon z,
\]
with $z\in[0,\bar{z}]$,  $\mp$ again referring to the counterclockwise, resp.\ clockwise beam,
and $\alpha$, $\bar{z}$ resp.\ equal to $\alpha_{\circlearrowleft}$, $\frac{c}{\bar{r}}\bar{t}_{\circlearrowleft}$
or $\alpha_{\circlearrowright}$, $\frac{c}{\bar{r}}\bar{t}_{\circlearrowright}$.
 As above, substitution 
in $g_{t\phi}/g_{tt}\cdot d\phi/dz$,  expansion in $\epsilon$,
integration in $dz$ form $0$ to $\bar{z}$ and multiplication by four allows to compute the first two integrals appearing
in (\ref{Sr}) as
\begin{equation}\label{I1+I2}
-\oint_{C_{\circlearrowleft}}\frac{g_{t\phi}}{g_{tt}}d\phi
-\oint_{C_{\circlearrowright}}\frac{g_{t\phi}}{g_{tt}}d\phi
=\frac{8\bar{r}^2\omega}{c}
\left[
1-\frac{1}{3}\left(\frac{\omega\bar{r}}{c}\right)^2-\frac{11}{30}\left(\frac{\omega\bar{r}}{c}\right)^4+...
\right].
\end{equation}
Analogously, by substituting in
$\sqrt{\frac{\gamma_{rr}}{|g_{tt}|}\frac{d^2r}{dz^2}+\frac{\gamma_{\phi\phi}}{|g_{tt}|}\frac{d^2\phi}{dz^2}}$,
expanding in $\epsilon$, etc.,
we evaluate the third and fourth integrals as
\begin{equation}\label{I3-I4}
\oint_{C_{\circlearrowleft}}\frac{1}{\sqrt{|g_{tt}|}}dl
-\oint_{C_{\circlearrowright}}\frac{1}{\sqrt{|g_{tt}|}}dl=
\frac{8\bar{r}^2\omega}{c}
\left[
-\frac{4}{15}\left(\frac{\omega\bar{r}}{c}\right)^4+...
\right].
\end{equation}
The first two terms  of (\ref{DtS}) correspond to the leading and next-to-leading order terms of (\ref{I1+I2}),
while higher order terms are correctly reproduced only when (\ref{I3-I4}) is added. In accordance with the revised formula
(\ref{Sr}), the sum of (\ref{I1+I2}) plus (\ref{I3-I4})  predicts the correct coordinate-time shift (\ref{DtS})
and consequently the correct proper-time shift (\ref{DtauS}) for the rotating square
interferometer.
This is due to the different paths followed by the counter propagating beams being correctly taken into account.
\end{description}

\begin{figure}[t]
 \centering
 \includegraphics[width=10.5cm]{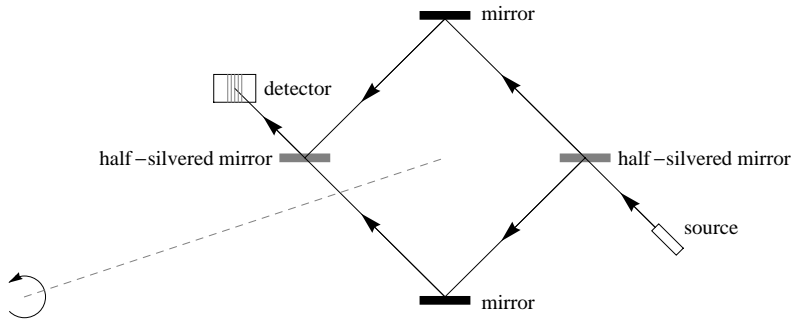}
\caption{A square Mach-Zehnder interferometer rotating around a point different from its center.
In presence of rotation the two counter-propagating beams can be re-directed to the detector
by adjusting the orientation of the two mirrors.}
 \label{Mach-Zehnder}
\end{figure}

In order to estimate relativistic corrections in a more realistic situation, we consider a square interferometer
rotating with angular speed $\omega$ around a point different from its center.
This might be an interferometer located on Earth's surface on the plane parallel to the
equatorial plane. It should be noted, that due to the lack of symmetry there is no longer guarantee that the paths followed
by the two counter propagating rays coincide at each mirror. Correspondingly, the two Sagnac loops no longer exactly
close as in figures \ref{inertial} and \ref{non inertial}. This possible source of error can be avoided
by replacing the Sagnac interferometer with a Mach-Zehnder interferometer. The two components of the original beam
follow then independent equal length  paths that can be separately tuned (see Fig.\ \ref{Mach-Zehnder}).
The time delay is half of the one measured in a Sagnac interferometer. Its analytical value is again given
by (\ref{Sr}), divided by two. The first two integrals combine now in the synchronization gap $\oint g_{0i}/g_{00} dx^i$, computed along the
union of the two paths effectively followed by light beams. The third and fourth integral still provide a small but non-vanishing
contribution extraneous to the synchronization gap. By proceeding as above, we obtain the coordinate-time delay
\begin{equation}\label{DtS'}
c\Delta t=\frac{4\bar{r}^2\omega}{c}
\left[
1+\left(\frac{R^2}{\bar{r}^2}-\frac{1}{3}\right)\left(\frac{\omega\bar{r}}{c}\right)^2+...
\right],
\end{equation}
with $R$ the distance of the interferometer form the center of rotation and $\sqrt{2}\bar{r}$ its side.
Assuming $r_{det.}\simeq R$ the corresponding proper-time delay is then given by
\begin{equation}\label{DtauS'}
c \Delta\tau=\frac{4\bar{r}^2\omega}{c}
\left[
1+\left(\frac{R^2}{2\bar{r}^2}-\frac{1}{3}\right)\left(\frac{\omega\bar{r}}{c}\right)^2+...
\right].
\end{equation}
Whereas the leading  Sagnac time delay does not depend on the position of the center of rotation, relativistic
corrections do. On the one side this has the effect of amplifying the order of the first correction from
$\left({\omega\bar{r}}/{c}\right)^2$  to the generally much larger $\left(\omega R/c\right)^2$.
For Earth's rotation $R\simeq 6.4 \times 10^{4}\textrm{m}$ and $\omega\simeq 0.7\times 10^{-4}\textrm{rad}/\textrm{s}$.
The detection of relativistic corrections requires therefore a relative sensitivity
$\Delta\omega/\omega=\frac{1}{2}\left(\frac{\omega R}{c}\right)^2\simeq 1.1\times 10^{-12}$.
Still four order of magnitude from the relative sensitivity of $10^{-8}$ of the large ring laser G in Wettzell \cite{Wet11},
but just one order of magnitude from the promised value of $10^{-11}$ of the multi-ring-laser gyroscope designed for the  measurement of the Lense-Thirring drag at 1\% \cite{B&al11}. On the other side, the dependence on $R$ opens to the theoretical possibility of a
direct measure of the radius of rotation. This would allow the realization of an absolute (non barometric)
altimeter, a tool of extraordinary importance in navigation systems and geophysical applications.

\section{Discussion}
\label{sec:3}
We have shown that the correct treatment of deflection of light due to rotational inertial forces,
requires a slight revision of the general relativistic Sagnac formula. The corrections involved are
too small ---second and higher order in the rotation parameter--- to be of practical
interest in present applications. However, their correct evaluation is important from a theoretical point of view.
In this respect a few observations are in order.\\
In discussing relativistic corrections to the Sagnac formula (\ref{S1913}) most authors consider the propagation of light
along a given curved path. In general a circular path.
The underlying idea is that any curved path can be approximated with arbitrary accuracy  by a polygonal path, realized for example
by an optical fiber.
While this limiting procedure makes perfect sense in mathematics, it necessarily stops at wavelength scales
in the physical reality. Moreover, the wave propagation in a curved wave-guide is only effectively free. Extrinsic curvature, torsion
and other constraint corrections \cite{Bur&al05} are in general much larger than the tiny relativistic corrections.
As a consequence equation (2) is not suitable for the computation of  relativistic corrections to the leading Sagnac term,
neither for loop interferometers, as proven in this paper, nor for fiber-optic gyroscopes. Correspondingly the time delay
in a Sagnac interferometer does not exactly correspond to the delay accumulated by a clock transported along an arbitrary closed
path described by (\ref{S}) \cite{AVS98,Min03}.
The crucial difference is that in the latter case light signals
are assumed to travel back and forth infinitesimally closed points: one first takes  the difference
between $(dt/dz)_{+}$ and $(dt/dz)_{-}$ and then integrates over the path. In the Sagnac effect instead, light travels a finite path:
one first integrates (over different paths) and then takes the difference of the results.\\
Next, we would like to comment on the analogy between the Sagnac effect in general relativity and the Aharanov-Bohm
effect in quantum mechanics \cite{AsMa75,Sak80,Ana81,HN90,RR03}.
As it is well known \cite{Anandan89,HN90}, the Sagnac effect as described by (\ref{S}) is not a fully topological
effect as the Aharanov-Bohm effect. This because the Coriolis-force field associated to the vector potential
$A_i=g_{0i}/g_{00}$ is in general non-vanishing. Nonetheless, the way the circulation of the vector potential
affects the wave phase in the two effects is identical.
However, our analysis shows that $\oint A_{i}dx^i$ only describes the leading contribution to the time shift in a Saganc
interferometer. The exact expression no longer writes as the circulation of a vector field. The contribution of
the third and fourth integrals in (\ref{Sr}) is in general very small, but not vanishing. Correspondingly,
the analogy between the Aharanov-Bohm effect and the phase shift in a Sagnac interferometer is broken by relativistic corrections.\\
Finally, we  remark that the considerations presented in this paper regarding the optical Sagnac effect,
equally apply to the Sagnac effect for matter beams propagating with a subluminal speed $v$
\cite{Pag75,Ana77,DY79,WSC79,RKW91,HN93,STK11}.
As for light beams, geodesics followed by matter particles traveling in opposite directions in a rotating frame are different.
This effect needs to be taken into account in equal footing.  Moreover, since the difference between clockwise and counter-clockwise
propagating paths increases by decreasing the particles speed, one should expect larger corrections. On dimensional
grounds, besides terms proportional to ${\omega^3}/{c^4}$ (see Eq. (\ref{DtauS'})), one should expect
extra terms going as  ${\omega^3}/{v^2c^2}$ and ${\omega^3}/{v^4}$. Preliminary
computations confirm this expectation. These terms are larger than the corrections for light beams. For
sufficiently slow particles they might become comparable and even larger than the classical Sagnac time shift.

\begin{acknowledgements}
We wold like to thank J. K. Pachos  for carefully reading the manuscript and related comments.
\end{acknowledgements}

\end{document}